\documentstyle[prd,aps,epsfig,eqsecnum,amssymb,amsmath,
                  floats,twocolumn]{revtex}
\tighten
\begin{document}
\clearpage
\preprint{}
\draft
\title{Prompt muon contribution to the flux underwater}
\author{ T. S. Sinegovskaya\,${}^{1}$ and S. I. Sinegovsky\,${}^2$
   }
\address{${}^1$Applied Physics Institute, Irkutsk State University,
             Irkutsk, 664003, Russia \\
       ${}^{2}$Physics Faculty, Irkutsk State University, Irkutsk,
             664003, Russia \\
    }
\setcounter{page}{1}
\maketitle
\begin{abstract}

 We present high energy spectra and zenith-angle distributions
 of the  atmospheric muons computed for the depths of the locations
 of the underwater neutrino telescopes. We compare the calculations
 with the data obtained in the Baikal and the  AMANDA muon experiments.
 The prompt muon contribution to the muon flux underwater due to
 recent perturbative QCD-based models of the charm production
 is expected to be observable at the depths of the large underwater
 neutrino telescopes. This appears to  be probable even at rather
 shallow depths (1-2 km), provided the energy threshold
 for muon detection is raised above $\sim  100$~TeV.
\end{abstract}

\pagenumbering{arabic}
\setcounter{page}{1}
\protect\section{Introduction}
\label{sec:Int}

Considerable literature exists on estimating the contribution to
cosmic ray muon fluxes that arises from the decay of charmed hadrons
\cite{VZ83,EGS83,BNS85,IKK86,VFGS,BNSZ89,TIG96,prd98,PRS99,MNS99,GGV99}.
Current data on the high-energy atmospheric muon flux obtained with many
surface and underground detectors are too conflicting  to provide
the means of probing charm production models (see, for example, \cite{prd98}).

 Both direct and indirect measurements of the atmospheric muon flux
 at sea level are limited to $\sim 70$~TeV for the vertical and to
 $\sim 50$~TeV for the horizontal.
 Statistical reliability of these data is still insufficient to
 evaluate the prompt muon contribution to the high-energy muon
 flux. Available energies and the accuracy of underground measurements
 are constrained because of the restricted size of detectors  and
 the uncertainties in the local rock density.
 Deep-sea installations have substantial advantages just due to large
 detector volume and homogeneous matter.
 So it is relevant to discuss the potential of the large
 underwater neutrino detectors (AMANDA, Baikal), in  the context of
 the prompt muon study, in future high-energy muon experiments.

 In this paper, we present calculations on zenith angle dependence of
 the high energy underwater muon flux taking into consideration
 the prompt muon fraction obtained in one of the recent perturbative
 QCD models of Pasquali et al. \cite{PRS99} (hereafter pQCD)
 in which small-x behavior of the gluon distributions is probed.
 This pQCD calculation based  on MRSD-~\cite{MRSD} and CTEQ3~\cite{CTEQ3}
 parton distribution functions (PDFs) includes the next-to-leading order
 (NLO) corrections  to the charm production cross sections.

 Perturbative QCD models differ in the PDF sets  being employed in the NLO
 calculations and in the choice of renormalization and factorization scales.
 A dependence on these quantities of the vertical  sea-level prompt lepton
 fluxes was  studied in Refs.~\cite{PRS99,GGV99,GGV3}.
 The predictions of the pQCD model~\cite{PRS99} are comparable to those of
 the earlier quark-gluon string  model~\cite{KP86} and the recombination
 quark-parton one (see~\cite{BNSZ89,prd98}).
 The muon spectra underwater obtained with the pQCD models and other types
 of the charm production models, the quark-gluon string  model (QGSM) and
 the recombination quark-parton model (RQPM), were partly discussed in
 Ref.~\cite{MNS99}.
 Here we would like to focus on variations in the expected
 underwater muon fluxes caused by distinctions between the PDFs used.
 In addition, we compare the expected underwater muon flux
 to the zenith angle distributions measured with the Baikal~\cite{NT-36}
 and the AMANDA~\cite{Amanda99} neutrino telescopes.

\protect\section{Sea-level muon fluxes}
\label{sec:SLMuons}

The atmospheric muon energy spectra and zenith angle distributions
of the conventional ($\pi,K$) muons, and the RQPM  and the QGSM
contribution, have been computed using the same nuclear cascade
model~\cite{prd98,VNS86,NSS98}. Let us glance over its key assumptions.

(i) The all-particle primary spectra and chemical composition
are taken according to Ref.~\cite{Nikolsky84}.
Nuclei of the primary cosmic rays are treated as the
composition of free nucleons.

(ii) Feynman scaling is assumed to be valid for hadron
produced in collisions of hadrons with nuclei of the atmosphere.

(iii)  The inelastic cross sections $\sigma_{hA}^{\it{inel}}$ for
       interactions of a hadron  $h$ ($= p, n, \pi^\pm$)
       with a nucleus $A$ grow logarithmically with the energy:
\[
\sigma_{hA}^{\it{inel}}(E_h) =
\sigma^0_{hA}+\sigma_A\ln(E_h/1\text{TeV}).
\]

(iv) The hadron production in kaon-nucleus and in charmed
     hadron-nucleus collisions is neglected.

(v) Charged pion is considered  stable in the kinetic stage of the
     nuclear cascade (not in the stage of muon production, to be sure).

(vi) Three-particle semileptonic kaon decays $K_{\mu3}$ are taken into account.

The energy spectrum of the conventional muons at the sea level
calculated for the vertical can be approximated (see~\cite{prd98})
for energies $E_\mu \gtrsim 1$\,GeV by the formula:

\begin{equation}\nonumber
{\cal D}_\mu^{\pi,K}\left(E_\mu,0^\circ \right) =
\begin{cases}
A_1 E_\mu^{-f(y)}, & \text{$E_1 \leq E_\mu\leq E_2$}\\
A_2 E_\mu^{-(1.791+0.304\, y)}, & \text{$E_2 \leq E_\mu\leq E_3$}\\
A_3 E_\mu^{-3.672}, & \text {$E_3\leq E_\mu \leq E_4 $} \\
A_4 E_\mu^{-4}, & \text{$E_\mu > E_4 $}.
\end{cases} \end{equation}
Here $E_\mu$ in GeV, $y=\log_{10}(E_\mu/1\text{GeV}),$
$$f(y)=0.3061+1.2743y-0.263y^2+0.0252y^3;$$
$A_1=2.95\times 10^{-3}, A_2=1.781\times 10^{-2},
A_3=14.35, A_4=10^3\,$ (cm$^{-2}$s$^{-1}$sr$^{-1}$GeV$^{-1}$);
$E_1=1$\,GeV,  $E_2=928$\,GeV,
$E_3=1587.8$\,GeV,  $E_4=4.1625\times10^5$\,GeV.

\begin{figure}[t!]
\vskip -5mm
\centering{\mbox{\epsfig{file=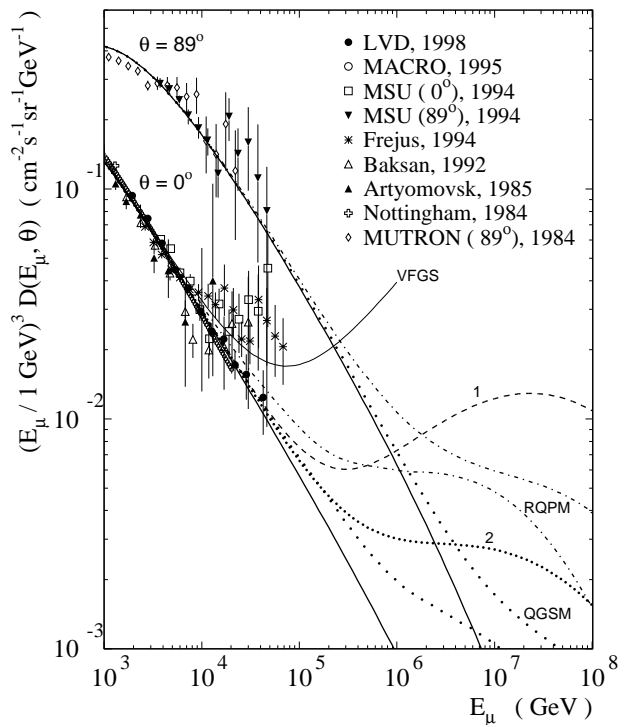,height=10.cm}}}
\vskip 2mm
\caption{Sea-level muon fluxes for the vertical and horizontal.
        The solid lines are for the conventional muons alone.
        Also shown are the conventional muons plus the
        prompt muon contribution estimated with several models:
        the pQCD-1 (dashed) and pQCD-2 (short dotted) for the  vertical;
        the model of Volkova et al. (thin) for the  vertical;
        the RQPM (dot-dashed) and QGSM (dotted) for the  vertical (lower)
        and near  the horizontal (up).
\label{fig-1}}
\end{figure}

 The results of the calculations of the muon zenith-angle
 distributions  at sea level are presented in Table~\ref{t1}
 for high energies 1-100 TeV.
The differential energy spectra (scaled by $E_\mu ^3$)
of the conventional muons at sea level are shown (solid) in
Fig.~\ref{fig-1} for the vertical and near horizontal direction
together with the data of the Nottingham spectrograph~\cite{Rastin84}
(one point at $E_\mu \simeq 1.3$ TeV), the MUTRON spectrometer~\cite{Mutron84},
and indirect measurements~\cite{MSU94,Frejus94,LVD98,ASD85,BNO92,MACRO95}.
\begin{table}[b!]
\protect\caption{Ratio ${\cal D}_\mu^{\pi,K}(E_\mu, \theta)/
{\cal D}_\mu^{\pi,K}(E_\mu, 0^\circ)$ of differential
 energy spectra of the conventional muons at sea level
 as a function of $sec\theta$.
\label{t1}}
\center{\begin{tabular}{c|ccccccc}
 $\sec\theta$ &\multicolumn{7}{c} {$E_\mu$ (TeV)}\\\cline{2-8}
      &1   &3    &5    &10   &30   &50   &100\\\hline
1.0   &1.0 &1.0  &1.0  &1.0  &1.0  &1.0  &1.0 \\
2.0   &1.74&1.86 &1.90 &1.93 &1.96 &1.96 &1.97 \\
3.0   &2.28&2.58 &2.67 &2.75 &2.82 &2.83 &2.84  \\
4.0   &2.66&3.12 &3.27 &3.40 &3.52 &3.54 &3.57  \\
5.0   &2.94&3.56 &3.76 &3.95 &4.12 &4.15 &4.19 \\
10.0  &3.53&4.69 &5.09 &5.50 &5.86 &5.95 &6.01  \\
15.0  &3.61&5.00 &5.49 &5.99 &6.45 &6.56 &6.65 \\
20.0  &3.57&5.05 &5.58 &6.12 &6.63 &6.75 &6.85 \\
40.0  &3.31&4.88 &5.44 &6.02 &6.56 &6.69 &6.79  \\
57.3  &3.17&4.74 &5.30 &5.88 &6.41 &6.54 &6.64 \\ 
\end{tabular}}
\end{table}
Open circles represent the MACRO best fit for the verical direction~\cite{MACRO95}.
(The detailed comparison between the calculated muon energy spectra
for different zenith angles and the sea-level experimental data,
in particular for large zenith angles, as well as calculations
of other authors, is made  in Ref.~\cite{Tanya99}.)
Dashed and dotted lines in Fig.~\ref{fig-1} correspond to the vertical
muon flux including the prompt muon contributions calculated \cite{PRS99}
with the CTEQ3 functions (pQCD-2) and the MRSD- set (pQCD-1).
Line 1 (dashed) corresponds to  the MRSD- set,  line 2 (short dotted)
corresponds to the  CTEQ3 PDF, both with factorization and renormalization
scales  $\mu_F=2\mu_R=2m_c$, and  with the charm quark mass $m_c=$ 1.3 GeV.
As evident from the figure, the pQCD predictions depend strongly on the PDF.

For comparison there are also shown predictions of
the charm production model of Volkova et al. (VFGS) \cite{VFGS}
(thin) and the results obtained with the RQPM (dot-dashed) and the
QGSM (dotted), both for the vertical direction
(lower) and near the horizontal (up). These results enable one
to make out the range of prompt muon flux predictions that
overspread more than one order of magnitude at $E_\mu \sim 1$ PeV.
It is interesting to note that old QGSM predictions~\cite{BNSZ89,KP86}
coincide practically with those of the pQCD-2 up to $\sim 600$~TeV,
while the RQPM flux appears to be close to the pQCD-1 one.

As is seen from Fig.~\ref{fig-1}, at $E_\mu \gtrsim 20$~TeV none
of the above models but the VFGS is consistent
with the data of MSU~\cite{MSU94} and Frejus~\cite{Frejus94}. Conversely,
none of the charm production models under discussion contradict
the LVD data~\cite{LVD98,LVD99}.
The VFGS, differing from the others in the extent of optimism,
gives the greatest prompt muon flux that is scarcely compatible
with the LVD upper limit~\cite{LVD99}.

The ``crossing energy'' $E_\mu^c(\theta)$ (the energy around
which the fluxes of conventional and prompt muons become equal)
depends on the choice of the PDF set.
The vertical crossing energy $E_\mu^c (0^\circ)$
is about 200 TeV for the pQCD-1 model, which is close
to the RQPM prediction ($E_\mu^c \sim 150$~TeV).
The vertical prompt muon flux predicted with the pQCD-2 model becomes
dominant over the conventional one at the energies $E_\mu \gtrsim 500$~TeV.
Therefore, in order for the differences between the pQCD models
to be found experimentally one needs to measure muon energies
above $\sim 100-200$~TeV.

\protect\section{Muon fluxes underwater}
\label{sec:UWMuons}

Muon energy spectra and zenith angle distributions deep
underwater are calculated using an analytical method~\cite{NSB94}
(see also Ref.~\cite{prd98}). By this method one can solve the
problem of the muon passing through dense matter for an arbitrary
ground-level muon spectrum and real energy dependence of differential
cross sections of muon-matter interactions.
The collision integral on the right-hand side of the muon transport
equation  describes the ``discrete''  energy loss of muons due to
bremsstrahlung, direct $e^+e^-$ pair production and photonuclear
interactions.

In this paper the ionization energy loss and the part of the loss
due to $e^+e^-$ pair production with $v < 2\cdot10^{-4}$
($v$ is the fraction of the energy lost by the  muon) is treated
as a continuous one: that is, the corresponding item is subtracted  from
the collision integral and transferred to the left-hand side
as a partial derivative with respect to energy
of the mean energy loss rate multiplied by the muon flux.

The calculations of the prompt muon fluxes underwater at different
zenith angles were performed with the parameterization of the
sea-level muon differential spectra (pQCD-1,2) taken from~\cite{PRS99}.

Omitting details, we dwell on a factor that may be useful in correcting
the underwater muon flux, provided that it is crudely estimated
with the continuous energy loss approximation (see Ref.~\cite{NSB94}).
This factor is the ratio $R_{d/c}$ of the integral muon flux
$I_\mu^{\rm disc}(E_\mu, h, \theta)$, computed for discrete
stochastic) muon energy losses, to the flux
$I_\mu^{\rm cont}(E_\mu,h, \theta)$
estimated with the continuous loss approximation.
\begin{table}[b!]
\protect\caption{Ratio $R_{d/c}=I_\mu^{\rm disc}/I_\mu^{\rm cont}$
 at $E_\mu > 10$~GeV.
\label{t2}}
\center{\begin{tabular}{c|c|cccc}
$\theta$ & $\sec\theta$ &\multicolumn{4}{c} {$h$ (km w.e.)}\\\cline{3-6}
(degrees)& &1  &2    &3   &4 \\\hline
$0   $ &1.0
 &1.02&1.05&1.09&1.15 \\
$60  $ &2.0           &1.04&1.14&1.31&1.58  \\
$70.53$ &3.0          &1.08&1.30&1.74&2.54  \\
$75.52$ &4.0          &1.12&1.55&2.53&4.79 \\
$78.46$ &5.0          &1.20&1.96&4.07&10.7 \\
$80.40$ &6.0          &1.30&2.60&7.21&28.7 \\
$81.79$ &7.0          &1.43&3.57&13.8&89.5  \\
$82.82$ &8.0          &1.58&5.00&28.7&284  \\
$83.62$ &9.0          &1.74&7.10&63.5&769  \\
$84.26$ &10.0         &1.92&10.5&151 &2320  \\ 
\end{tabular}}
\end{table}
%

In Table~\ref{t2} the ratio $R_{d/c}$ is given as a function of the
water depth and zenith angle for muon energies above $10$~GeV.
As is seen the effect of discrete energy loss for the large depth
is far from being small:
 $R_{d/c}$ is about 2 for the depth value of $\sim 10$~km w.~e.
The ratio is slightly affected by zenith-angular dependence
of the sea-level muon flux.
More precisely, the $R_{d/c}$ depends both on the  ``spectral index''
of the muon flux and geometric factor of $\sec \theta$.
The former varies weakly with zenith angle while the latter
plays more important role in the $R_{d/c}$  defining
the thickness of water layer  $X=h \sec \theta$ that a muon overpasses
($h$ indicates the vertical depth in km).

For water the ratio $R_{d/c}$ as a function of the slant depth $X$
can be approximated with accuracy better than $\sim 10\% $ as
\begin{eqnarray} 
 R_{d/c}&=&0.99+0.02X+6.74\cdot10^{-4}X^3, \ \, \,  X=1-12~{\rm km};
                 \nonumber    \\
 R_{d/c}&=&1.43+0.054\exp[(X-1.19)/3.64], \ \,  X=12-35~{\rm  km}.
 \nonumber \end{eqnarray}

The effect of the discrete loss increases as the muon energy grows.
 The energy dependence of the ratio $R_{d/c}$ is adequately illustrated
 by  the following: for the depth of 12 km w.~e. $R_{d/c} \simeq $2.5
 at $E_\mu=$10~GeV and $R_{d/c} \simeq $4.0 at $E_\mu= 1$~TeV.

\begin{figure}[t!]
\centering{\mbox{\epsfig{file=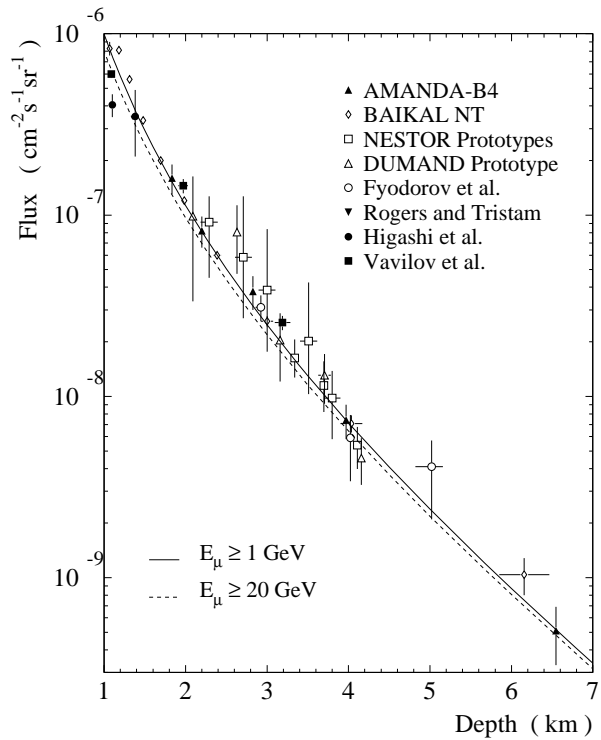,height=10.cm}}}
\vskip -0mm
\caption{Vertical muon flux as a function of water depth.
        The lines correspond to the $\pi,K$-muons
        calculated with the muon residual energy
        above 1 GeV (solid) and above 20 GeV (dashed).
\label{fig-dir}}
\end{figure} 

In Fig.~\ref{fig-dir} we present a comparison between the expected
muon vertical depth--intensity relation in water and the data
obtained in underwater experiments
(see for review~\cite{prd98},\cite{NT-36}), including recent
measurements in the AMANDA-B4 experiment~\cite{Amanda99}.
The computation was performed with water parameters:
$\varrho=1\, g/cm^3$,\, $<Z>=7.47$,\, $<A>=14.87$,\, $<Z/A>=0.5525$,\,
$<Z^2/A>=3.77$. The muon energy loss per $1\,g/cm^2$ in ice is
considered to be equal that in water but $\varrho_{ice}=0.92\,g/cm^3$.
The calculations are presented for the muon residual energy
(threshold of the detection) $E_\mu \geq $1~GeV (solid) and
$E_\mu\geq $20~GeV (dashed). This difference needs to be considered
especially for shallow depth.

\begin{figure}[t!] 
\centering{\mbox{\epsfig{file=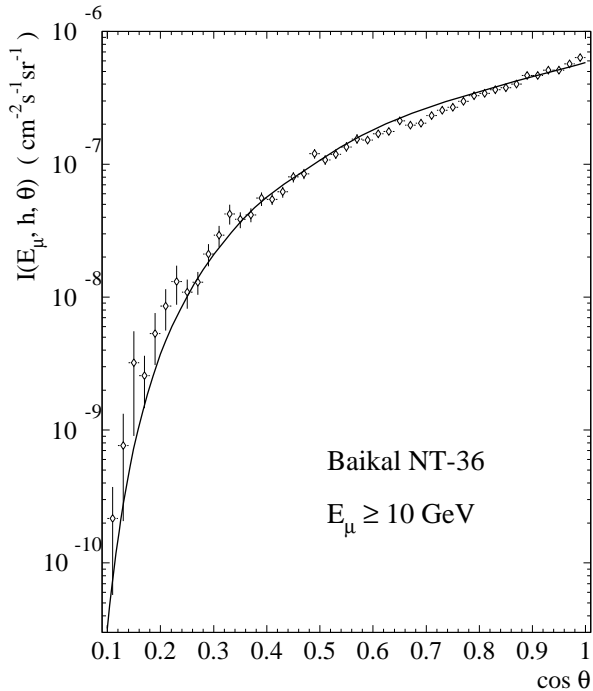,height=10.cm}}}
\vskip -0mm
\caption{Zenith angle distribution of the muon flux underwater
 measured by Baikal NT-36~\protect\cite{NT-36}.
\label{fig-nt36}}
\end{figure}
\begin{figure}[t!]
\centering{\mbox{\epsfig{file=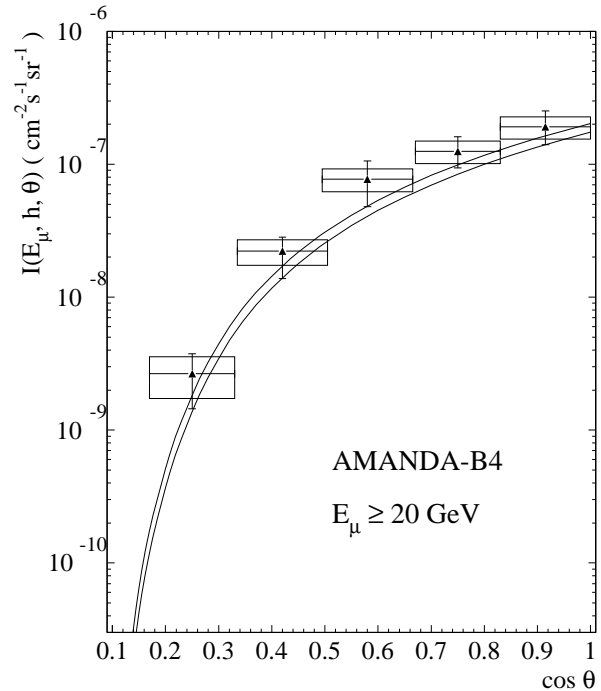,height=10.0cm}}}
\vskip -0mm
\caption{Zenith angle distributions of the muon flux underwater
  measured with the AMANDA-B4~\protect\cite{Amanda99}.
\label{fig-amanda}}
\end{figure}

Figs.~\ref{fig-nt36},\,\,\ref{fig-amanda} show a comparison
of the predicted muon zenith angle distribution
(without considering the prompt muon contribution) with the
measurements in the neutrino telescopes  NT-36~\cite{NT-36} and
AMANDA~\cite{Amanda99}.

The line in Fig.~\ref{fig-nt36} presents the
calculation for the muon threshold energy
$E_\mu= $10~GeV at depth $h=$1.15~km.  Our calculation is in
reasonable agreement with the measurements of the NT-36
at all  but the angle range $80-84^\circ$.
In Fig.~\ref{fig-amanda}, the upper line relates to the flux at the depth
$h=$1.60~km w.~e.  calculated for the muon residual energy
$E_\mu \geq $20~GeV, the lower one relates to $h=$1.68~km w.~e.
for the same energy threshold.
The difference illustrates the possible effect of an uncertainty
in determining the average ``trigger depth''~\cite{Amanda99}
(relating to the center of gravity of all hit optical modules
 in the AMANDA-B4 experiment).
The computed angle distribution agrees fairly well
with the AMANDA-B4 data including zenith angles $\theta > 70^\circ$.

\begin{figure}[b!]
\centering{\mbox{\epsfig{file=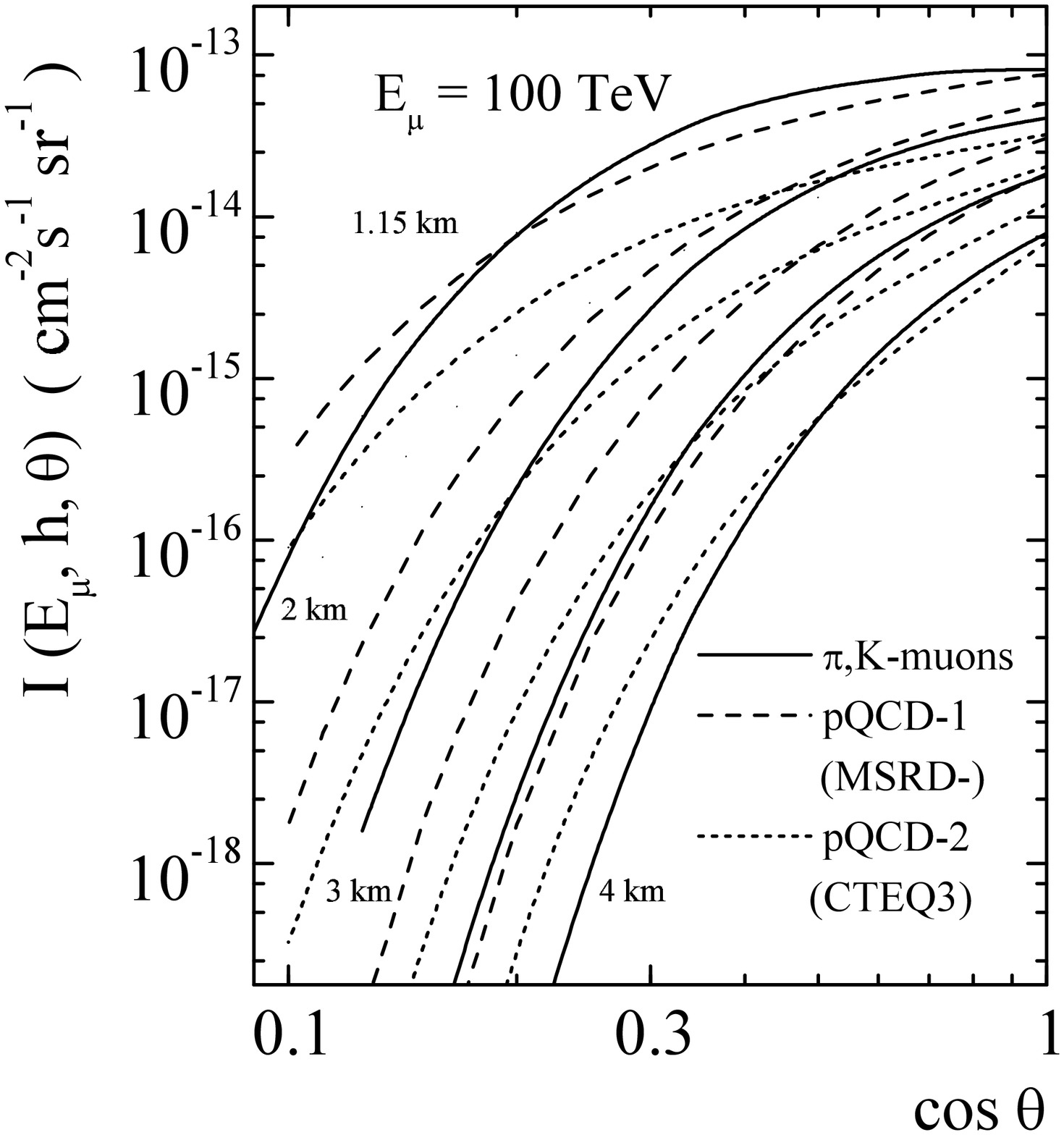,height=10.0cm}}}
\vskip -15mm
\caption{Fluxes of muons above 100~TeV at water depth
 $h=1.15$, $2$, $3$, $4$~km (from top to bottom)
 as a function of cosine of the zenith  angle.
\label{fig-5}}
\end{figure} 

The contributions of the ($\pi,K$) and prompt muons underwater
to zenith angle distribution at $E_\mu>100$~TeV calculated for
four values of depths (of 1.15 to 4~km) are shown in Fig.~\ref{fig-5}.
Here we present the results obtained with the pQCD-1 (dashed) and
the pQCD-2 (dotted). It is interesting to note that  the dashed
line representing the pQCD-1 prompt muon contribution twice
intersects the line of the conventional
flux at $h=1.15$~km: near the vertical and at $\theta\sim 75^\circ$.
This can occur because of the different zenith angle
dependence of the conventional muon flux and the prompt muon one.
And this means that at a depth of 1.15 km the nearly
doubled muon event rate (for $E_\mu>100$~TeV) would be observed in
the $0-75^\circ$ range, instead of the rate expected due to
conventional muons alone.

There is no intersection of the pQCD-2 line  at  $h=1.15$~km up to
$\theta\sim 85^\circ$. The intersection point shifts to smaller
zenith angles with increasing depth. For a depth of 2 km (nearly
the AMANDA depth) it is possible to observe prompt muon fluxes
that would be expected with the pQCD-2 at not too large angles
($\sim 70^\circ$).
It should be mentioned that the underwater prompt muon flux
will be distorted in a large zenith angle region because the
angle isotropy approximation considered for the predictions of the
pQCD models  is valid only at $\theta\lesssim70^\circ$ and
$E_\mu\lesssim10^3$~TeV.

\begin{figure}[t!]
\centering{\mbox{\epsfig{file=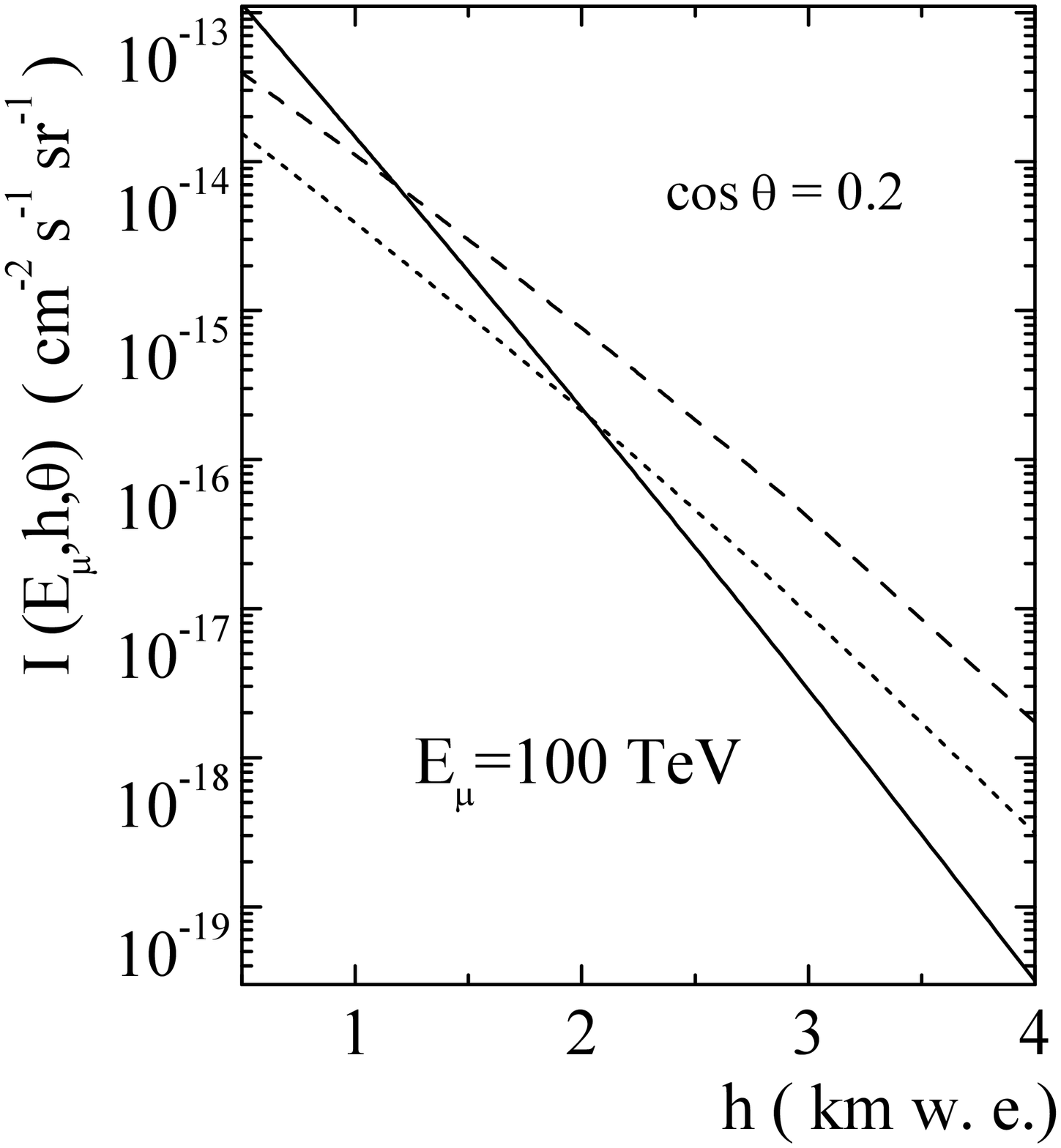,height=10.0cm}}}
\vskip -15mm
\caption{ The muon flux underwater against depth
 at $\cos~\theta =0.2$.
 The contributions shown are the conventional muons (solid) and the
 prompt muons due to the pQCD-1 (dashed) and the pQCD-2 model (dotted).
\label{fig-6}}
\end{figure} 

The depth dependence of the muon flux underwater at zenith angle
of $\sim 78^\circ$ (Fig.~\ref{fig-6}) indicates that in the case of
the pQCD-1 one can observe the doubling of the muon flux at the
Baikal depth of 1.15 km for $E_\mu \geq 100$~TeV.
At a depth $\sim 2$~km the same takes place even with the lesser
prompt muon flux predicted with the pQCD-2 model.

\begin{figure}[t!]
\centering{\mbox{\epsfig{file=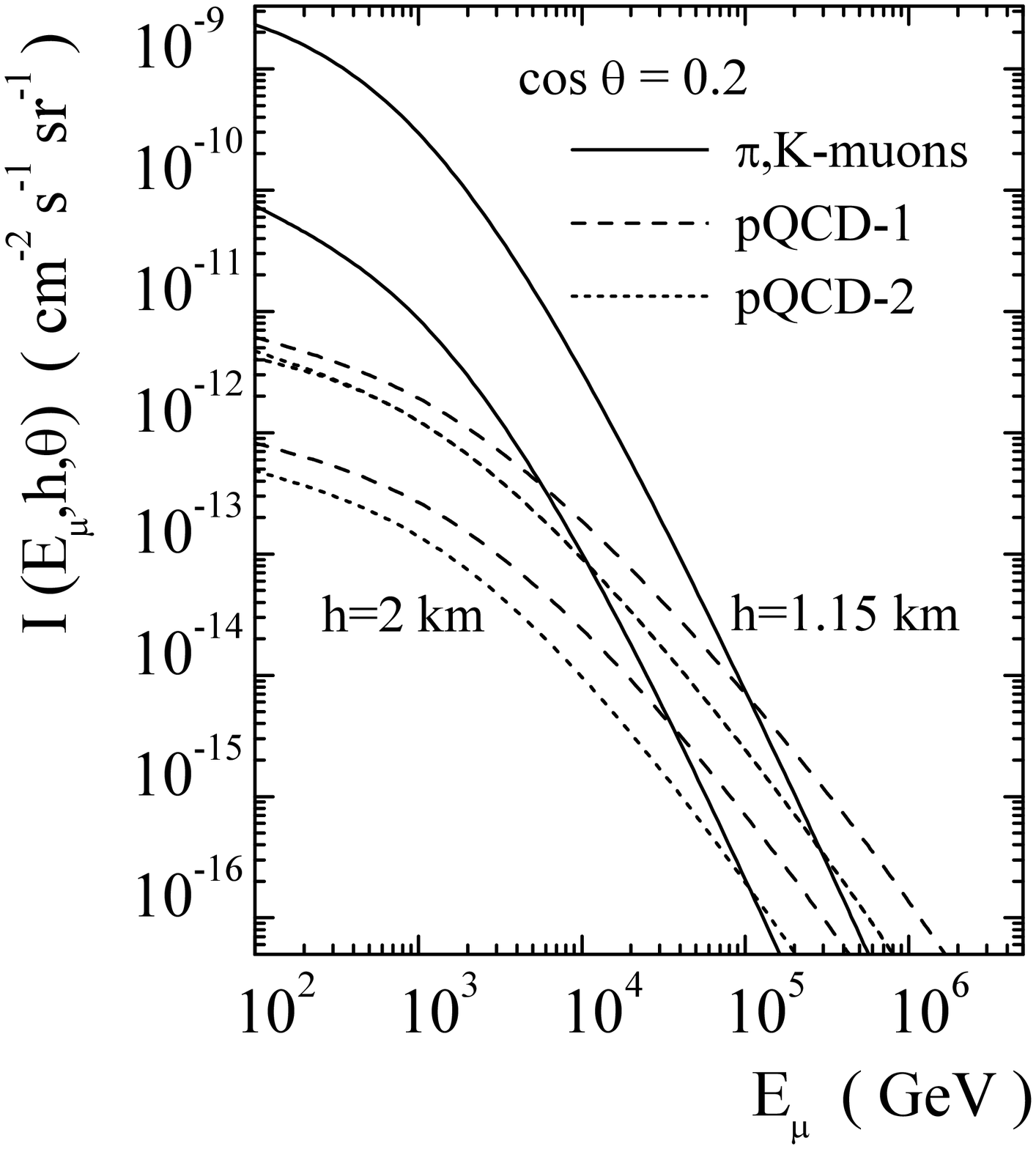,height=10.0cm}}}
\vskip -15mm
\caption{Integral muon spectra underwater at zenith angle
 $\theta=78.5^\circ$ at a depth of  1.15 km (upper) and of
 2 km (lower).
The contributions shown are the conventional muons (solid)
and the prompt muons (dashed and dotted).
\label{fig-7}}
\end{figure}

Fig.~\ref{fig-7} shows muon integral energy spectra at a depth
$h=1.15$~km (Baikal) and 2~km (AMANDA) and for
$\cos\theta=0.2$ ($\theta\simeq78.5^\circ$).
Also presented are the predictions of the prompt muon flux issued
from the pQCD-1 (dashed) and  pQCD-2 (dotted).  The crossing energies
$E_\mu^c (\theta)$ at the AMANDA depth are less than  the ones at the
Baikal depth by a factor of $\sim 3$.
In particular, the PQCD-1 model gives
$E_\mu^c(\theta\simeq78.5^\circ)\approx30$~TeV at $h=2$ km
and  $E_\mu^c(78.5^\circ)\approx100$~TeV at a depth of $1.15$ km.
The same quantity calculated with pQCD-2 is 100 and 250 TeV
respectively.

One can see (Fig.~\ref{fig-7}) that the AMANDA depth ($\sim 2$~km)
gives, in a sense, the definite advantage in comparison with
the Baikal one. Indeed, in the former case the assumed threshold
energy is less, the muon flux difference between the pQCD-1 model
and the pQCD-2 one is larger (up to two orders of magnitude),
and the expected event rate remains approximately equal to the rate
at the Baikal depth.

It should be pointed out that muon residual energies below
$\sim 10$~TeV and zenith angle $\theta\lesssim 75^\circ$
would be available (see Ref.~\cite{MNS99} for a discussion),
in the above context, in future high-energy muon
experiments with the NESTOR deep-sea detector~\cite{NESTOR}
which is expected to deploy at a depth of about 4 km.

\protect\section{Summary}
\label{sec:Concl}

Energy spectra and zenith angle distributions of the atmospheric muons
at high energies have been calculated for the depths
from 1 to 4~km that correspond to the depths
of operation of large underwater neutrino telescopes. The estimation
of the  prompt muon contribution performed with the pQCD-1,\,2
shows that the crossing energy $E_\mu^c$ above which the prompt muon flux
becomes dominant over the conventional one, is within the range
of $\sim 200$ to $\sim 500$~TeV at sea-level, depending on the
choice of the parton distribution functions.
For the flux underwater at a zenith angle $\sim 78^{\circ}$,
the pQCD-1 model leads to the value $E_\mu^c \simeq 30$~TeV ($h=2$~km)
and $E_\mu^c \simeq 100$~TeV ($h=1.15$~km). The corresponding crossing
energies for the pQCD-2 model are $E_\mu^c \simeq 100$
and $E_\mu^c\simeq 250$~TeV.

The absolute value of the muon flux underwater around $E_\mu^c$
depends on the charm production model. This circumstance enables,
in principle, bounds to be put on the charm production cross section
based on  measurements of zenith angle distributions of the muon flux
at high energies.
In particular,  PDF sets under discussion, the MRSD- and the CTEQ3,
differ in the small-$x$ behavior of the gluon distribution:
$ xg(x)\sim x^{-\lambda} $, where $\lambda\simeq 0.29 \div 0.35$
for the CTEQ3 against $\lambda=0.5$ for the  MRSD- set.
(See Ref.~\cite{GGV3} for the $\lambda$-dependence of the sea-level
prompt muon flux.)
These PDFs yield inclusive cross sections of charmed
particles produced in nucleon-air collisions and charm production cross
sections that diverge rapidly from each other with increasing energy.
For muon energies above 100 TeV and for $\cos\theta=0.2$
these differences lead to the fact that prompt muon flux predicted with
the pQCD-1 exceeds the flux arising from the pQCD-2 model
by a factor of about 4 at $h=1.15$~km or about 5  at $h=2$~km.

In conclusion we outline three probable ways for solving
of the prompt  muon  problem in the underwater experiments.
First, one can measure zenith angle dependence of the muon flux in the
energy region of $50-100$~TeV (see Fig.~\ref{fig-5}): the expected
event rate with the Baikal NT-200 is about $200-300$ per year per
steradian, supposing that the effective area of NT-200 is  $10^4$~m$^2$
for  $E_\mu \geq 100$ TeV~\cite{NT-200}.

Second, the flux with  muon energies $E_\mu \geq 100$~TeV
measured as a function of depth (say, in depth region about
$0.8-1.2$~km) at a given  zenith angle ($\sim 78^\circ$),  could enable
the charm production models to be discriminated (see Fig.~\ref{fig-6})
at the event rate level of about 200 per year per steradian.

And last, one can attempt to extract information on the prompt muon
flux underwater from muon integral spectra being measured at a given
depth and at a given zenith angle (Fig.~\ref{fig-7}). In this case
the event rate is a factor 7 less than in the previous one (with the
NT-200 capabilities).
It should be pointed out that the AMANDA
depth of $\sim 2$~km  provides some advantage: the threshold
energy is less, the muon flux difference between the pQCD-1 model
prediction and the pQCD-2 one is greater, and the expected event rate
remains approximately equal to that at the Baikal depth.

\acknowledgments

We thank V.~A.~Naumov for helpful discussions and
comments, S.~Hundertmark and C.~Spiering for kindly
providing the table data on the muon zenith-angle distribution and
the depth-intensity relation measured in the AMANDA-B4 experiment.

 This work is supported in part by the Ministry
 of Education of the Russian Federation under grant No.~015.02.01.04
 (the Program ``Universities of Russia -- Basic Research'').

\end{document}